\documentclass[a4paper,jcp,aip,twocolumn, 10pt]{revtex4-1}

\usepackage{graphicx}
\usepackage{amsmath,amssymb,amsfonts}
\usepackage{bm,natbib}
\usepackage{subfigure}
\usepackage{color}
\usepackage[latin1]{inputenc}

\begin{document}

\title[Dipolar filtered magic-sandwich-echoes as a tool for probing molecular motions using time domain NMR]
{Dipolar filtered magic-sandwich-echoes as a tool for probing molecular motions using time domain NMR}

\author{Jefferson G. Filgueiras} 
\affiliation{Instituto de Física de São Carlos, Universidade de São Paulo, P.O. Box 369, São Carlos, 13560-970 SP, Brazil}

\author{Uilson B. da Silva} 
\affiliation{Instituto de Física de São Carlos, Universidade de São Paulo, P.O. Box 369, São Carlos, 13560-970 SP, Brazil}

\author{Giovanni Paro}
\affiliation{Instituto de Física de São Carlos, Universidade de São Paulo, P.O. Box 369, São Carlos, 13560-970 SP, Brazil}

\author{Marcel N. d'Eurydice} 
\affiliation{School of Chemical and Physical Sciences, Victoria University of Wellington, PO Box 600, Wellington, New Zealand}

\author{Márcio F. Cobo}
\affiliation{Instituto de Física de São Carlos, Universidade de São Paulo, P.O. Box 369, São Carlos, 13560-970 SP, Brazil}

\author{Eduardo R. deAzevedo} 
\affiliation{Instituto de Física de São Carlos, Universidade de São Paulo, P.O. Box 369, São Carlos, 13560-970 SP, Brazil}

\begin{abstract}
We present a simple $^{1}$H NMR approach for characterizing intermediate to fast regime molecular motions using $^1$H time-domain NMR at 
low magnetic field. The method is based on a Goldmann Shen dipolar filter (DF) followed by a Mixed Magic Sandwich Echo (MSE). The dipolar 
filter suppresses the signals arising from molecular segments presenting sub kHz mobility, so only signals from mobile segments are 
detected. Thus, the temperature dependence of the signal intensities directly evidences the onset of molecular motions with rates 
higher than kHz. The DF-MSE signal intensity is described by an analytical function based on the Anderson Weiss theory, from where parameters 
related to the molecular motion (e.g. correlation times and activation energy) can be estimated when performing experiments as function of 
the temperature. Furthermore, we propose the use of the Tikhonov regularization for estimating the width of the distribution of correlation 
times.
\end{abstract}

\maketitle

\section{Introduction}

The development of advanced materials with desired properties and function requires a thorough knowledge of the structural and dynamical properties 
of the molecular systems that comprise them. Many properties of synthetic and natural organic materials are closely related to the mobility 
of their molecules, making molecular motions a key point for understanding these systems \cite{HernandezGluten,FariaPFO,fariajpcb2009,HansenAdvMat,Kurz,HongJPCB,DupreeNature,andronis1998}. In this respect, Nuclear Magnetic 
Resonance (NMR) provides a variety of methods to access motion information in a wide frequency range (Hz-MHz) \cite{SpiessAnniversary,HansenRev,Krushelnitsky,deAzevedoPNMRS}. 

For solid systems, the NMR methods developed for probing molecular motions usually rely on the  orientation dependence of the NMR frequencies, 
which provides high sensitivity to the time scale and geometry of segmental motions \cite{HansenRev,Krushelnitsky,deAzevedoPNMRS}. Such features can be studied with 
unpaired precision by using samples with isotopic labeling, for instance $^{2}$H and $^{13}$C, at specific molecular segments. This approach, 
however, becomes uneconomical in studies of large molecules. When isotopic labeling is not possible (or desired), several high resolution 
solid-state NMR methods have been developed\cite{HansenRev,Krushelnitsky,deAzevedoPNMRS,deAzevedo2009}, giving site-resolved information about the molecular dynamics. However, they lack sensitivity due to the detection of low abundant nuclei, as $^{13}$C, resulting in time consuming experiments.  In contrast, $^1$H based solid-state NMR methods provide high sensitivity, but loses in specificity, since the local character of the dipolar interaction is lost due to 
spin-diffusion \cite{rothwelljcp1981,deazevedo2000,palmer2001,faske2008,cobopccp2009,lange2011}. Despite not being able to deliver details on the local geometry, $^{1}$H solid-state NMR is useful to study the overall dynamical behavior and its temperature dependence, estimating the activation parameters of the motion.

Because of the high sensitivity of $^{1}$H detection this experiments can be efficiently performed at low magnetic fields 
using much less expensive and simpler NMR spectrometers. Because of that, Time Domain $^1$H NMR (TDNMR) at low magnetic field, became quite 
popular for studying soft matter, in particular polymer and biopolymer based systems \cite{Schaler2013,Saalwachter2012,Papon2011,refMSEmotion}. There are many TDNMR techniques devoted to study molecular motions exploring either the relaxation phenomena \cite{rothwelljcp1981,palmer2001,lange2011} or the motional averaging of the $^{1}$H-$^{1}$H dipolar interaction and its effect on the NMR signals\cite{Schaler2013,Saalwachter2012,Papon2011,refMSEmotion}.

Here we address the effects of motions on common TDNMR pulse sequences and propose an approach where a Goldmann-Shen type of dipolar filter \cite{GS1}, with duration t$_{f}$, is followed by a mixed Magic Sandwich Echo sequence (DF-MSE)\cite{Matsui}. The dipolar filter suppresses the signal of $^{1}$H in "rigid" segments (i.e. which the time-domain signal decays faster than the filter time), keeping the $^{1}$H signal arising from more mobile segments. Thus, at low temperatures where all molecular segments are rigid, the signal will be completely suppressed and the intensity of the MSE echo will vanish. However, in a temperature where some molecular segments become mobile, the dipolar filter sequence no longer suppresses the signal and the MSE echo 
appears. Thus, the method will detect the onset temperature of specific molecular motions by the increase of the DF-MSE echo intensity 
as a function of temperature.  
   
We also demonstrate that the filter time dependence of the normalized DF-MSE intensities acquired at a fixed temperature is sensitive to both 
the mean and the width of the distribution of correlation times. A simple formula based on the Anderson-Weiss approach \cite{AWeiss} is then 
used as a kernel for Tikhonov  regularization procedure providing a fit to the experimental data that allow to estimate the correlation time distribution in am model independent fashion. Using this procedure we obtain the distributions of the correlation times and activation parameters of motions associate to the glass transition of amorphous polymers and compare our results with those obtained by other methods \cite{WhitePIB, Zemke}. All measurements were carried 
out in a time considerably shorter in comparison to these other techniques.

The article is presented as follows. Section II is devoted to a short theoretical background, reviewing some key concepts. Section III discusses the results, presenting the method to detect the molecular motions along with some numerical simulations and experimental results on the characterization of the molecular motion in model amorphous polymers. At last, some discussions and conclusions are presented.

\section{Theory}

\subsection{Description of the DF-MSE pulse sequence}

In this section we discuss an approach for the detection of molecular motions in the intermediate to fast frequency range (kHz-MHz) using a dipolar filter. In principle, any type of dipolar filter could be used \cite{SpiessDF,MAPE}, but we choose a slightly modified version of the 
simpler and well known Goldmann-Shen (GS) dipolar filter \cite{GS1}. The most known application of GS filter is to suppress 
the signal of rigid segments in spin diffusion experiments, which aim to characterize domains sizes and interphases in heterogeneous  polymers, like 
semicrystalline polymers, copolymers and polymer blends \cite{MAPE,SpinDiffBlumich_KaySchaeller,Schaler2013}.

\begin{figure}
\centering
\includegraphics[width=\columnwidth]{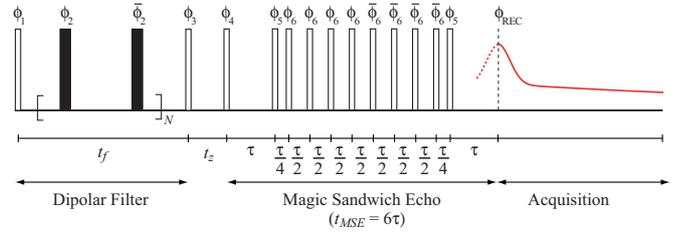}
\caption{The DFMSE pulse sequence, the signal after the GS dipolar filter comes solely from the slow-decaying components, the MSE pulse 
train is employed to overcome the experimental dead time. The filled rectangles indicate the $\pi$ pulses, while all the other
pulses are $\pi/2$ rotations. The phase cycling is given in Table \ref{pc}.}
\label{fig:sequence}
\end{figure}

The basic GS pulse sequence consists of three pulses, see Fig. \ref{fig:sequence} ignoring the $\pi$ pulses. The first pulse leads the magnetization to the $xy$-plane. During a waiting time $t_f$, called filter time, the magnetization due to rigid segments rapidly decays due to the strong 
$^{1}$H-$^{1}$H dipolar coupling. In contrast,  the magnetization from mobile segments, i.e which have the $^{1}$H-$^{1}$H dipolar 
coupling averaged by motions, remains after $t_f$ and one of its component ($x$ or $y$ in consecutive scans) is stored in the $z$-direction 
by the second pulse. The remaining component in the $xy$-plane after the second pulse is removed  after $t_z$ by proper phase cycling of the pulses. 
Thus, the magnetization after $t_z$ is along the $z$-direction. Because the signal from strong dipolar coupled $^{1}$H spins decay during $t_{f}$, the signal after the GS filter originates only in segments where the $^{1}$H-$^{1}$H dipolar coupling is reduced by molecular motions. Thus, $t_f$ defines the upper limit for the motional rates that will be interpreted as rigid in the experiments. 

\begin{table}[t]
 \caption{Phase cycling for the DFMSE pulse sequence.}
 \label{pc}
 \centering
  \begin{tabular}{cccccccc}  \hline\hline
   Step ~~~& $\phi_1$ ~~~& $\phi_2$ ~~~& $\phi_3$ ~~~& $\phi_4$ ~~~& $\phi_5$ ~~~& $\phi_6$ ~~~& $\phi_{REC}$ \\ \hline
   $1$  ~~~& $0$      ~~~&  $90$    ~~~& $180$    ~~~& $0$      ~~~& $270$    ~~~& $0$      ~~~& $180$     \\
   $2$  ~~~& $0$      ~~~&  $270$   ~~~& $0$      ~~~& $0$      ~~~& $90$     ~~~& $0$      ~~~& $180$     \\
   $3$  ~~~& $90$     ~~~&  $0$     ~~~& $270$    ~~~& $0$      ~~~& $90$     ~~~& $0$      ~~~& $180$     \\
   $4$  ~~~& $90$     ~~~&  $180$   ~~~& $90$     ~~~& $0$      ~~~& $270$    ~~~& $0$      ~~~& $180$     \\
   $5$  ~~~& $180$    ~~~&  $90$    ~~~& $0$      ~~~& $0$      ~~~& $0$      ~~~& $270$    ~~~& $0$       \\
   $6$  ~~~& $180$    ~~~&  $270$   ~~~& $180$    ~~~& $0$      ~~~& $180$    ~~~& $270$    ~~~& $0$       \\
   $7$  ~~~& $270$    ~~~&  $0$     ~~~& $90$     ~~~& $0$      ~~~& $180$    ~~~& $270$    ~~~& $0$       \\
   $8$  ~~~& $270$    ~~~&  $180$   ~~~& $270$    ~~~& $0$      ~~~& $0$      ~~~& $270$    ~~~& $0$       \\
   \hline\hline
  \end{tabular}
\end{table}

When using the GS filter to probe molecular motion one should also consider segments that do not move fast enough to fully average the 
$^{1}$H-$^{1}$H dipolar couplings. In this case the evolution under residual $^{1}$H-$^{1}$H dipolar couplings during the GS sequence 
can generate multiple quantum coherences (MQCs) during $t_z$. These MQCs can be converted into transverse magnetization by the third 
pulse, leading to artifacts in the detected signal \cite{Packer}. In the present case, the single and double quantum coherence existing 
during $t_z$ have their effects canceled by cycling the phases of the first two pulses, with respect to the third, in steps of 
$90^o$ \cite{Wokaun}.

Another point to be considered concerns the static magnetic field inhomogeneity. For highly inhomogeneous magnetic fields, as typical in TDNMR, 
the decay of the signal during $t_f$ is dictated by the magnetic field inhomogeneity. As a result, the upper limit for $t_f$  would be defined by the field inhomogeneity. To overcome it we added a train of $\pi$ pulses \cite{HanhEcho_CPMG} during $t_f$, see Fig. \ref{fig:sequence}. We remark that 
such $\pi$ pulses have no effect on the spins evolution under the $^{1}$H-$^{1}$H dipolar coupling if the pulse lengths are much shorter 
than the inverse of the dipolar coupling strength.

The signal from strong dipolar coupled $^{1}$H spins decays in few microseconds, so it is necessary to avoid significant loss of magnetization due 
to finite receiver dead time. This is achieved using a mixed-MSE of duration $t_{MSE}$  before detection \cite{Matsui}. The  mixed-MSE sequence, displayed in the Fig.\ref{fig:sequence}, provides the refocusing of both the $^{1}$H-$^{1}$H dipolar coupling and the field inhomogeneities. 

The intensity of the echo after applying the GS filter followed by the mixed-MSE pulse sequence is then
\begin{eqnarray}
I_{DF}(t_f,t_{MSE}) = G(t_f)I_{MSE}(t_{MSE}),
\end{eqnarray}
\par\noindent where $G(t_f)$ and $I_{MSE}(t_{MSE})$ are, respectively, the functions defining the modulation in the original 
magnetization introduced by the application of the GS filter and the mixed-MSE sequence.

\subsection{Detecting onset temperatures of molecular motions using the DF-MSE pulse sequence}

The qualitative behavior of $I_{DF}(t_f,t_{MSE})$ offers information about the onset temperature of molecular motions in a very simple way, which
can be quite useful for many multiphase systems, where  the macroscopic properties are associated to molecular dynamics in different segments. 
Therefore, it is important to investigate the onset temperatures of dynamic processes in distinct segments or phases. In principle, 
this can be achieved using such as Differential Scanning Calorimetry (DSC), Dynamical Mechanical Thermal Analysis 
(DMTA) or Dielectric Relaxation\cite{Boyd1,Boyd2}. However, multiphase systems have wide distribution of motions and complexes 
structures. Consequently,  DSC usually fails in detecting local motions or eve bulk processes in highly heterogeneous systems. Although
DMTA and Dielectric relaxation are both suitable methodologies, they may require special sample preparation, which is not always possible or 
easy. On the other hand, $^{1}$H NMR based methods have good sensitivity, but the clear identification of the onset temperatures usually 
requires some data processing and simulations \cite{deAzevedoPNMRS,palmer2001}. Therefore, our aim here is to provide a direct and very straightforward 
way to identify the onset temperature of molecular motions in TDNMR using the DF-MSE experiment. 
    
Our approach resembles previous works by Saalwächter and co-workers which used the measurement of the intensity of a mixed-MSE echo 
$I_{MSE}(t_{MSE})$ as a function of temperature. In principle, for rigid systems the  temperature dependence of $I_{MSE}(t_{MSE})$  
is dictated by the Currie law, i.e., proportional to T$^{-1}$. However, molecular motions occurring with rates in the order of the 
inverse of the $^{1}$H-$^{1}$H dipolar coupling interfere with the echo formation, reducing the signal intensity in a Curie law 
independent fashion. Thus, by monitoring the echo intensity as a function of temperature, the onset of molecular motions can be probed 
as the deviation from the Curie law at temperatures where the motion sets on. In reference \cite{Papon2011} it was shown that 
the normalization of the echo intensities by the Curie temperature dependence provides a clear way of detecting the onset temperature 
of molecular motions, such as glass transition \cite{Papon2011} or local flips of polymer chains\cite{KayKerstin}. It is worth mentioning 
that the signal intensity may vary as a function of temperature due to instrumental reasons, for example, the efficiency of the pulses 
and resonance offsets may change with temperature. 

Here we use the same principle, but with the DF-MSE pulse sequence. In this case the temperature variations of the signal due to factor 
unrelated to motion can be minimized by the normalization of $I_{DF}(t_f,t_{MSE})$ by the intensity of a signal acquired 
with short $t_f$,  so the GS dipolar filter have no effect. Supposing that the temperature variation 
of the signal by factors other than motion could be encoded in a parameter $\alpha(T)$, one can build the normalized intensity as 
    
\begin{eqnarray}\label{normsignal}
I_{DF}^{N}(t_f,T) &=& \frac{G(t_f,T)I_{MSE}(t_{MSE},T)\alpha(T)}{G(0,T)I_{MSE}(t_{MSE},T)\alpha(T)} \cr &=&  \frac{G(t_f,T)}{G(0,T)}.
\end{eqnarray}

\par\noindent Interestingly, the normalized intensity only depends on the effect of the GS dipolar filter on the signal intensity. The onset temperature of specific molecular motions will be detected by the increase of the DF-MSE echo intensity as a function of 
temperature, which will be seen as a change in the slope of the DF-MSE echo $I_{DF}^{N}(t_f,T)$ vs. $T$ curve.  
 
\subsection{Quantifying motion effects using DF-MSE}

Here we discuss the dependence of the signal $I_{DF}$ as a function of the parameters related to the molecular motion. First, 
we show how the correlation time affects the intensity of the detected echo as well as the width of the distribution of correlation times. 
The description in terms of the temperature is carried out using an Arrhenius activation function, but can 
be easily extended to other activation functions such as Vogel-Fulcher-Tammann (VFT)  or William-Landel Ferri (WLF)  \cite{activationfunctions}. 
So, we are able to check how parameters like the activation energy and the Arrhenius pre-factor affect the DF-MSE signal.  
    
As stated in Eq. (\ref{normsignal}), the normalized intensity $I_{DF}^{N}$ depends only on the motion effects encoded during the 
GS filter, i.e $G(t_f)$. This is a simple free induction decay (FID) during $t_f$, for  which the motion effects 
can be well predicted in the framework of the Anderson and Weiss approach \cite{AWeiss,Papon2011}.
    
According to the Anderson-Weiss approximation of a stationary correlation function \cite{AWeiss}, the FID is described by
\begin{eqnarray}\label{AWeiss}
G(t) = \exp\left(-M_{2}\int_{0}^{t}(t-\tau)K(\tau)d\tau\right),
\end{eqnarray}
\par\noindent
\par\noindent where $K(\tau)$ is the memory function, i.e., the normalized correlation function of the shifted frequencies due to the 
local fields, and satisfies $K(0) = 1$. $M_2$ is the second moment of the distribution of local 
fields in the absence of molecular motions.

Considering homonuclear spin systems, the local field distribution is defined by the $^{1}$H-$^{1}$H dipolar coupling, which has 
a Gaussian profile \cite{VanVleck,Hirschinger1,Hirschinger2}. Also assuming isotropic rotational diffusion of the 
molecules \cite{Kimmich1, Kimmich2, Reichert}, the memory function is given by\cite{AWeiss}
\begin{eqnarray}
K(\tau) = \exp\left(-\frac{|\tau|}{\tau_c}\right), \nonumber
\end{eqnarray}
\par\noindent
where $\tau_c$ is the correlation time of the motion. Within these approximations, the normalized FID signal is written as
\begin{eqnarray}\label{FIDsignal}
G(t) = \exp\left(-M_{2}\tau_{c}^{2}(\exp(-t/\tau_c) + t/\tau_c - 1)\right).
\end{eqnarray}
\par\noindent A more detailed discussion of the formalism is presented in \cite{Hirschinger1, Hirschinger2, Kimmich2}.

Therefore, the normalized intensity $I_{DF}^{N}= \frac{G(t_f,\tau_{c})}{G(0,\tau_{c})}$ becomes
\begin{eqnarray}\label{normsignalAW}
I_{DF}^{N}(t_f,\tau_{c}) =\exp\left(-M_{2}\tau_{c}^{2}(\exp(-t_f/\tau_c) + t_f/\tau_c - 1)\right).
\end{eqnarray}
\par\noindent Note that the temperature dependence in Eq. (\ref{normsignal}) is related to $\tau_c$ 
by the activation function of the motion.

\begin{figure}[htbp]
\centerline{\includegraphics[width=\columnwidth]{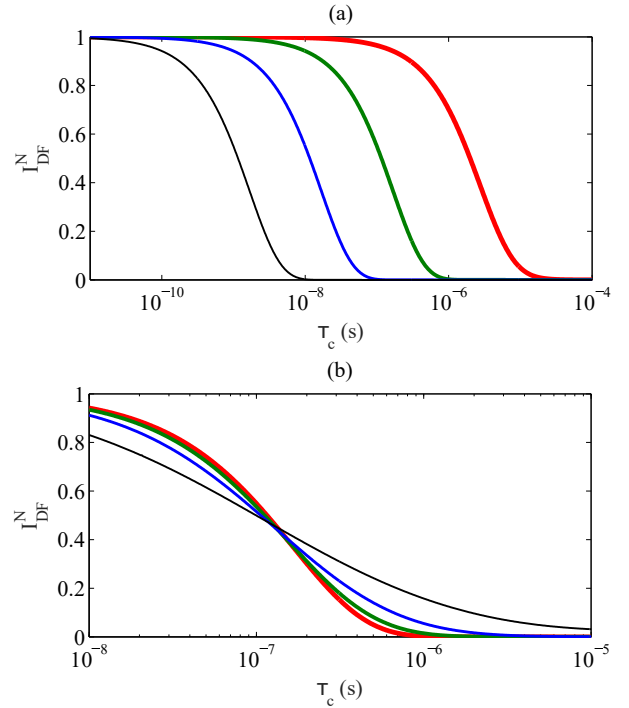}}
\caption{$I_{DF}^{N}$ as a function of correlation times. For different filter times, (i) shows the sensitivity of the 
dipolar filter. The simulations were carried out with a single correlation time, for $t_f =$  50 $\mu$s (red, thickest), 
1000 $\mu$s (green), 10000 $\mu$s (blue) and 100000 $\mu$s (black, thinnest). Below, we show the effects of a distribution of correlation 
times, assumed to be a log-gaussian distribution $g(\tau_c, \sigma, \left\langle\tau_c\right\rangle) = 
\exp(-\frac{(\ln \tau_c - \left\langle\ln\tau_c\right\rangle)^2}{2\sigma^2})/\sigma\sqrt{2}$ , where $\left\langle\tau_c\right\rangle$ is the average
correlation time and $\sigma$ is the standard deviation. For a fixed filter time $t_f =$ 1000 $\mu$s, 
we made calculations for $\sigma =$ 0.1 (red, thickest), 0.5 (green), 1.0 (blue) and 2.0 (black, thinnest). }
\label{fig:corr_time}
\end{figure}

Fig. \ref{fig:corr_time} (a) shows calculated $I_{DF}^{N}$ as a function of $\tau_c$, for different $t_f$ and $M_2= 6\times 10^{9}$ (rad/s)$^2$.
As the correlation time decreases the dipolar coupling starts to be averaged out by the molecular motions, leading to a remaining
magnetization after the dipolar filter. The amplitude of this magnetization increases until the correlation time is 
short enough to fully average the $^{1}$H-$^{1}$H dipolar coupling, so the filter has no effects on the magnetization 
during the filter time. We also observed that the range of correlation times that $I_{DF}^{N}$ is sensitive to the 
motion, i.e the dynamic window of the experiments, can be adjusted by setting the filter times 
$t_f$.  It is worth mentioning that the dynamical window is limited by the relaxation times $T_1$, as the signal starts 
to show relaxation effects during $t_f$ rather than simple averaging of the $^{1}$H-$^{1}$H dipolar coupling . Note that the effect of the random fluctuation of the dipolar fields, i.e. $T_{2}$, is taken into account in the AW theory.

The previous equation is restricted to the case of motions occurring with a single correlation time.  Since many systems have heterogeneous 
dynamics, one must consider the case where the motion occurs with a distribution of correlation times. This can be achieved by 
calculating the $I_{DF}^{N}$ intensity as a sum of the intensities given by Eq. (\ref{normsignalAW}) weighted by the  distribution of 
correlation times function $g(\tau_c)$, i.e, 

\begin{eqnarray}\label{FIDsignalw}
I_{DF}^{N}(t_f) = \int_{0}^{\infty}I_{DF}^{N}(t_f,\tau_{c})g(\tau_c)d\tau_c,
\end{eqnarray}
    
To demonstrate the effects of the distribution of correlation times on $I_{DF}$ we assume a log-gaussian distribution of 
correlation times \cite{log-gaussianSpiess1,log-gaussianSpiess2} which can be characterized by two parameters, the standard deviation 
of the distribution $\sigma$ and the center of the distribution in the log scale $\left\langle\ln\tau_c\right\rangle$, see caption 
of Fig. \ref{fig:corr_time}. Fig. \ref{fig:corr_time} (b) shows a set of curves calculated for $I_{DF}^{N}$  as a function of 
$\left\langle\tau_c\right\rangle$ for different $\sigma$ values. As it can be observed, the larger the value of $\sigma$, the wider 
the $I_{DF}^{N}$ vs. $\left\langle\tau_c\right\rangle$ curves. This occurs because the amount of segments 
moving inside a fixed dynamical sensibility window, defined by $t_f$, changes more slowly with increasing $\left\langle\tau_c\right\rangle$ 
as $\sigma$. 
    
\subsection{Temperature dependence of DF-MSE signals}\label{taucvsT}

To check the signal dependence on the temperature, we discuss how each of the parameters in the activation function modify the signal. 
For the sake of simplicity, only the Arrhenius function is considered in the calculations, but similar behaviors are observed when using 
other activation functions \cite{activationfunctions}. According to the Arrhenius activation function the mean correlation time can be 
converted into temperature using two activation parameters, the apparent activation energy $E_a$ and the prefactor $\tau_{0}$, i.e, 

\begin{equation}\label{tcTArrFct}
\ln\left(\frac{\langle\tau_c (T)\rangle}{\tau_0}\right) = \frac{E_a}{RT},
\end{equation}

\par
\begin{figure}[htbp]
\centerline{\includegraphics[width=\columnwidth]{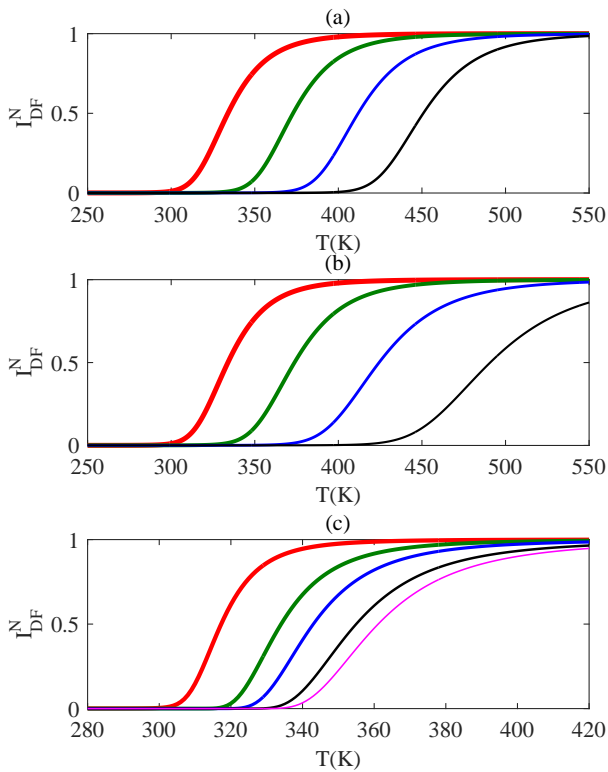}}
\caption{Signal dependence on temperature. Each curve in (i) is simulated for a different value of the activation energy, while 
the Arrhenius prefactor is kept constant at $\tau_0 = 10^{-15}$~s . $E_a$ takes the values 60 kJ/mol (red,thickest), 67 kJ/mol (green), 
74 kJ/mol (blue) and 81 kJ/mol (black, thinnest). (ii) exhibits how the Arrhenius prefactor affects $I_{DF}^{N}$. $E_a = 60$~kJ/mol 
for all curves. $\tau_0$ values are 10$^{-15}$~s (red, thickest), 10$^{-14}$~s (green), 10$^{-12}$~s (blue) and 10$^{-12}$~s (black, thinnest). 
For both (i) and (ii), the filter time is set to 50 $\mu$s. Simulations for different filter times are shown in (iii), with 
$E_a =$ 60 kJ/mol and $\tau_0 =$ 10$^{-15}$~s. The filter times are 50 $\mu$s (red, thickest), 250 $\mu$s (green), 800 $\mu$s (blue), 
2000 $\mu$s (back) and 3000 $\mu$s (magenta, thinnest). }
\label{fig:sim_filtertime_temp}
\end{figure}

\noindent $R$ is the gas constant. Fig. \ref{fig:sim_filtertime_temp} shows  $I_{DF}^{N}(t_f,T)$ vs. $T$ curves calculated with different 
values of the $E_a$ and $\tau_{0}$. The calculations were carried out using a filter time of $50~\mu$s, which is long enough to suppress 
the contribution of the signal from rigid segments.  In Fig. \ref{fig:sim_filtertime_temp}(a), the signal $I_{DF}^{N}(t_f,T)$ is calculated 
for different values of $E_{a}$. The curve is shifted to higher temperatures as $E_{a}$ increases, with a small decrease on its slope. 
If $\tau_{0}$ is changed, the dynamic window is also translated, with a more significant decrease in the curve slope, as can be seen in Fig. \ref{fig:sim_filtertime_temp}(b).

\subsection{DF-MSE difference curves}

So far, we discussed how the Goldman-Shen dipolar filter is suited to probe molecular motions and 
to estimate the activation parameters of such motions. It is important to exploit ways to quantify the 
kinectic parameters itself, i.e., the correlation times and their distributions. 

It is worth mentioning that the effects of a distribution of correlation times are hard to take into account, 
since a model for the distribution width as a function of the temperature is not known, to the best of our 
knowledge.  Wachowicz \emph{et al.} applied a linear model to describe $\sigma(T)$ \cite{WhitePIB,White2}, however, 
this is not a generally expected behavior. Indeed, more elaborated and time consuming $^{2}$H and $^{13}$C exchange NMR experiments have been used to extract the width assuming log-gaussian or Kohlrausch-Williams-Watts (KWW) correlation functions. Thus, a fast method capable to estimate the distribution of correlation times, without previous assumptions, is highly desirable. 

Fig. \ref{fig:sim_filtertime_temp} (c) shows the signal dependence on temperature for different filter times. 
As the filter time increases, the strenght  of the dipolar filter is increased, which  implies on the dynamical window moving 
to smaller correlation times, i.e., higher temperatures.  Hence, the choice of the filter time sets the cut-off 
motion rate detectable by the dipolar filter. Based on this idea one can take the differences between two 
$I_{DF}^{N}(t_f,\tau_{c})$ intensities with different filter times ($t_{f}^{min}$ and $t_{f}$) to obtain a well defined dynamic window. 

\begin{eqnarray}\label{defdiff}
I_{diff}(t_f,t_{f}^{min},\tau_c) = I_{DF}^{N}(t_{f}^{min},\tau_c) - I_{DF}^{N}(t_{f},\tau_c).
\end{eqnarray}
\par\noindent

Considering the $\tau_c$ dependence, $I_{diff}(t_f,t_{f}^{min},\tau_c)$ is null when the signals acquired with $t_{f}^{min}$ and $t_{f}$ 
are equal and nonzero whenever they are different. Therefore, the nonzero region of $I_{diff}(t_f,t_{f}^{min},\tau_c)$ defines the $\tau_c$ 
boundary of a well defined dynamic window. If the motion occurs with a distribution of correlation times $I_{diff}(t_f,t_{f}^{min},\tau_c)$ 
will depend on how many segments moves within the specified dynamic window, i.e. on the distribution of correlation times itself. 
To have the widest dynamic window possible, we choose the minimum filter time, $t_{f}^{min}$, as the shortest one without any signal from 
the rigid segments. It is worth mentioning that the quantity above keeps its dependence with the correlation times, just like 
$I_{DF}^{N}(t_{f},\tau_c)$. Besides, establishing a well defined dynamical window enhances the dependence of the distribution of correlation times.  Taking into account the motion heterogeneity of the system Eq.~\eqref{defdiff} becomes a sum over the 
correlation times weighted by the corresponding distribution $g(\tau_c)$, i.e,

\begin{equation}\label{IdiffDist}
I_{diff}(t_f,t_{f}^{min}) = \int_{0}^{\infty}I_{diff}(t_f,t_{f}^{min},\tau_c)g(\tau_{c})d\tau_c.
\end{equation}

\subsection{Extracting the distribution of corrrelation times from the filter time dependence of $I_{diff}$}

For fixed $\langle\tau_{c}\rangle$ and $t_{f}^{min}$ the dependence of $I_{diff}(t_f,t_{f}^{min})$ on $t_f$ encodes the distribution 
of correlation times. This quantity can be easily obtained experimentally by measuring a set of normalized DF-MSE echoes at fixed temperatures 
and with filter times ranging from $t_{f}^{min}$ to $t_{f}^{max}$. Assuming a specific shape for the distribution of correlation time, like a log-gaussian distribution, $I_{diff}(t_f,t_{f}^{min})$ vs. $t_f$ can be calculated for various $\langle\tau_{c}\rangle$ and $\sigma$. The fit of these calculated 
curves to the experimental data provides the values for $\langle\tau_{c}\rangle$ and $\sigma$. A similar procedure has been used with 
different NMR methods \cite{refMSEmotion,WhitePIB,White2,CODEX,PUREXinterm,log-gaussianSpiess2}. 
 
However, Eq.~\eqref{IdiffDist} is a Fredholm integral of first kind, which means that if the exact expression of $I_{diff}(t_f,t_{f}^{min})$ 
is known it can be used as a fitting Kernel in a Tikhonov regularization procedure. Such procedure allows us to extract $g(\tau_{c})$ without 
any previous assumption on the shape of the distribution. Interestingly, this can be done for several temperatures independently, so the 
temperature dependence of $g(\tau_{c})$ can be obtained. Moreover, once $g(\tau_{c})$ is known, it is straightforward to extract the 
$\langle\tau_{c}\rangle$, so the motion activation function can also be properly mapped. There are some drawbacks on this approach. 
First, the Anderson Weiss treatment considers isotropic rotation diffusion, so geometrically restricted motions are not covered. This can be relaxed 
choosing a proper AW formula adapted to restrict rotations. However, it requires some knowledge on the dynamic order parameter of such 
motions\cite{Hirschinger1, Hirschinger2, Dipshift1}. Second, the the ambiguities on the Tikhonov 
regularization, mainly considering the choice of the regularization parameter are well known \cite{Tikhonov}, so a good set of data, with 
enough $t_{f}$ points and minor experimental noise and artefacts, is required. Since we are dealing with $^{1}$H acquisition and 
the pulse sequence of Fig.~\ref{fig:sequence} is quite robust to experimental artefacts, we believe such approach is suitable.

\section{Experimental results}

\begin{figure}
\centering
\includegraphics[width=\columnwidth]{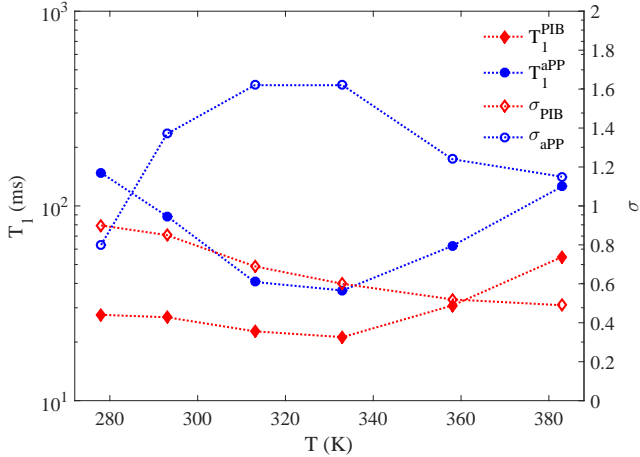}
\caption{Standard deviations of the distributions of correlation times and measured values of T$_1$. The red diamonds depict the 
PIB data, while blue circles exhibit the aPP data. The filled symbols are the standard deviations.}
\label{fig:sigma}
\end{figure}

\begin{figure*}[t]
\includegraphics[width=\textwidth]{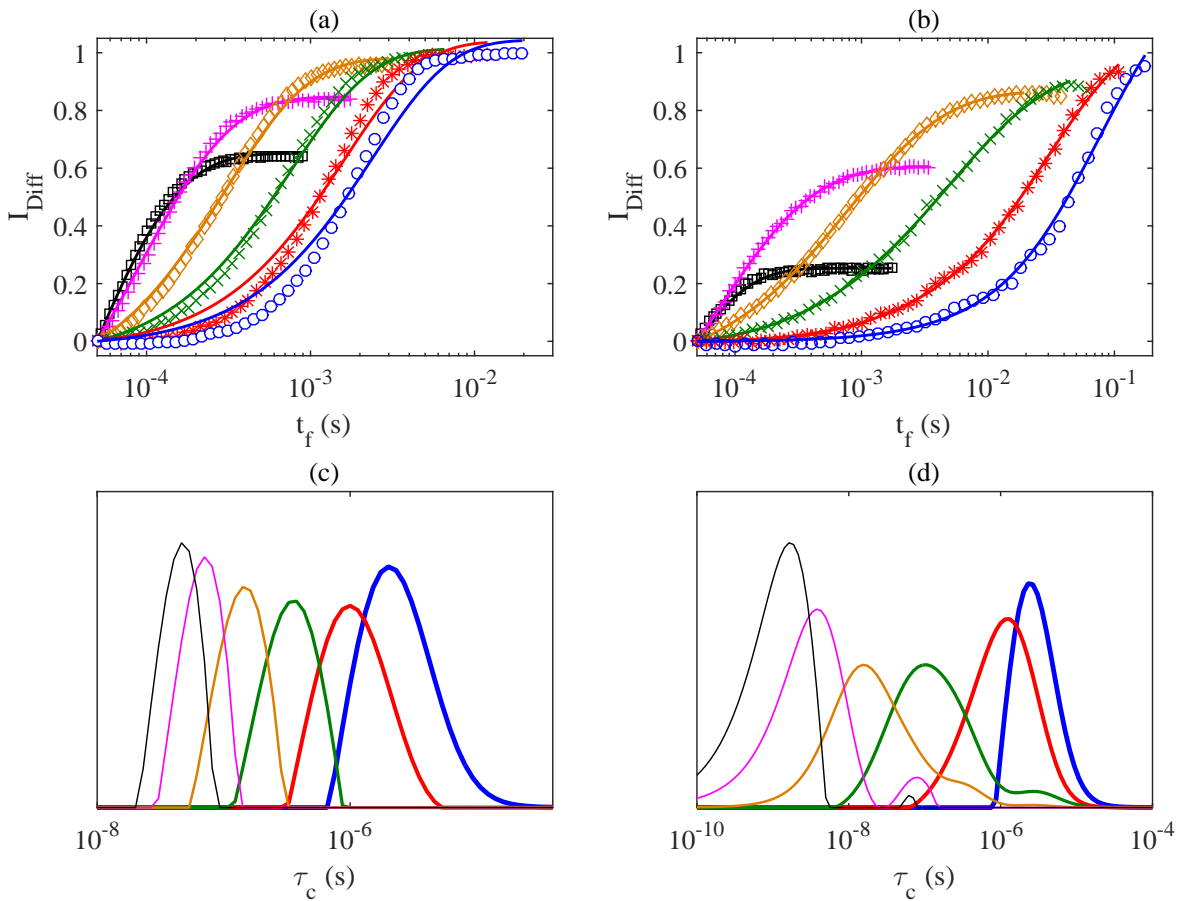}
\caption{Fittings and the respective distributions of correlation times for aPP and PIB samples. Above, (a) and (b) show the PIB experimental data 
and fittings for six different temperatures (T $=$ 278 K (black, squares), 293 K (magenta, $+$), 313 K (orange, diamonds), 333 K (green, $\times$), 
358 K (red, $*$) and 383 K (blue, circles)). Solid lines represent the fittings. The respective distributions obtained are shown in (c) 
and (d), with the same color scheme as the fittings and data, and line thickness decreasing with the temperature.}
\label{fig:fits}
\end{figure*}

\textbf{NMR}: measurements were performed on a 20 MHz Bruker MINISPEC. The $^{1}$H $\pi$ and $\frac{\pi}{2}$ pulse lengths 
were 2.5$\mu s$ and 5$\mu s$, respectively.  We measured T$_1$ using a standard inversion-recovery sequence, and we set the recycle 
delays to 5T$_1$. The Magic Sandwich Echo (MSE) sequence was performed with echo times of 100$\mu s$. Dipolar filtered MSE experiments 
were carried out using a Goldman-Shen filter sequence varying the filter times t$_{f}$.

\textbf{Samples}: The samples used in the experimental demonstrations were atactic poly(propylene): aPP [CH$_{2}$-CH(CH$_{3}$)-],
Mw = 118000 g/mol, T$_{g}$ = 251 K and poly(isobutylene): PIB [CH$_{2}$-C(CH$_{3}$)$_{2}$-], Mw = 2400000 g/mol, 
T$_{g}$ = 205 K. Both samples are fully amorphous. We measured the second moment for both samples fitting the FID, at 173 K,
with the Abragam function\cite{Abragam}. The values obtained were $(6.62 \pm 0.09)$ x$10^9$ (rad/s)$^2$ and $(8.96 \pm 0.03)$ x$10^9$ 
(rad/s)$^2$, for the aPP and PIB, respectively. More details about the samples can be found in references 
\cite{PUREXinterm, Zemke}. The temperature was controlled using a 
Bruker BVT3000, with a 2 K resolution on the sample.

\subsection{Estimating correlation times distributions}

Fig. \ref{fig:fits} shows the filter time dependence of experimental $I_{diff}$ intensities for PIB, Fig. \ref{fig:fits} (a), and
aPP, Fig. \ref{fig:fits} (b), at six different temperatures. In both cases we used $t_{f}^{min}=50\mu s$. There is a remarkable
displacement of the curves for longer $t_{f}$ values as the temperature increases, related to the decrease of
the mean correlation time of the motion. Moreover, considering that $I_{diff}$ represents the fraction of segments moving in a
dynamic window defined by $t_{f}^{min}=50\mu s$ and $t_{f}$, one should expect that $I_{diff}$ reaches $1$ only when all segments
are moving inside such window. Thus, the increasing in the plateau values of the curves with temperature suggests the
distribution of the correlation times getting narrower as the temperature increases. This narrowing is observed for the PIB sample,
after performing the Tikhonov regularization to fit the data, as can be seen in Figs. \ref{fig:sigma} and Fig. 
\ref{fig:fits} (c). To provide an evidence of such behavior we fitted each correlation time distribution of Fig. \ref{fig:fits} 
using log-gaussian functions. The standard deviations of the distributions, as a function of temperature, are shown in Fig. 
\ref{fig:sigma}, along to the measured values for T$_1$. The expected decay of the width of the distribution of correlation
times as a function of temperature is observed. There is an almost linear dependence of the standard deviation with the temperature, 
as assumed in \cite{WhitePIB}. However, the width of the distribution of correlation times for the aPP presents an unexpected maximum
around 333 K, as seen in Figs. \ref{fig:sigma} and \ref{fig:fits} (d). This distinct behavior of aPP and PIB is explained 
by comparing the $\sigma$ vs. $T$ and $T_1$ vs. $T$ curves, Figs. \ref{fig:sigma} and \ref{fig:fits}, in each case. As can 
be observed, for PIB $T_1\gg t_{f}$ in all the measured temperature range, so the effect of $T_1$ in the magnetization evolution 
during the dipolar filter can be neglected for all temperatures. However, for temperatures around the minimum of $T_1$ this is not 
true for aPP. Indeed, the profiles of $\sigma$ vs. $T$ and $T_1$ vs. $T$ are remarkably anti-correlated, suggesting that the 
apparent broadening of the aPP distribution of correlation times is related to the $T_1$ relaxation time, when $t_{f}\approx T_1$. 
This shows that the estimation of the correlation time distribution by DF-MSE  is only reliable if $T_1\gg t_{f}^{max}$, where 
$t_{f}^{max}$ is the longest filter time. All results were obtained without any \emph{a priori} 
assumption on how the standard deviation behaves with the temperature. 

Despite the drawback in the precise estimation of the distribution of correlation times width by DF-MSE, the mean values of 
this distribution might still be used for providing motion activation parameters. This relies on the observation that
$\langle\tau_{c}\rangle$ dictates the behavior of the $I_{Diff}$ curves for short $t_f$, when the T$_1$ effects are still negligible. 
Furthermore, the mean-values of the distributions obtained from the fittings are used to compare our data with previous experiments, 
using a William-Landeu-Ferry 
(WLF) activation function,
\begin{eqnarray}\label{WLF}
\log_{10}\left(\frac{\langle\tau_{c}(T)\rangle}{\tau_{c}(T_g)}\right) = -\frac{C_1(T - T_g)}{C_2 + T - T_g},
\end{eqnarray}
\par\noindent where $\langle\tau_c (T)\rangle$ and $\tau_c(T_g)$ are the correlation time at the temperature $T$ and at the glass transition $T_g$, respectively. 
$C_1$, $C_2$ and $\tau_c(T_g)$ are fitting parameters. 

\begin{figure}[htbp]
\centering
\includegraphics[width=\columnwidth]{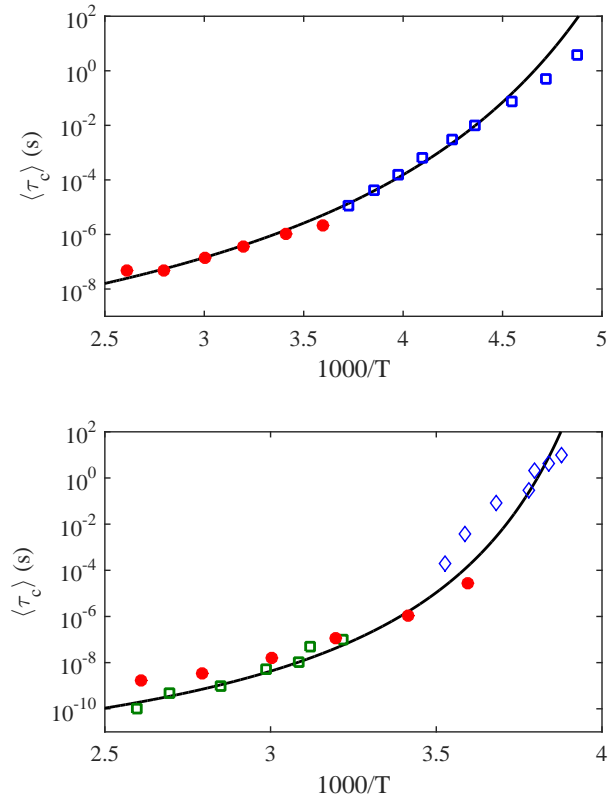}
\caption{Mean correlation time for the molecular dynamics of PIB and aPP. Above, the PIB data (red, circles), along with the data from 
$^{13}$C NMR exchange experiments\cite{WhitePIB} (blue, squares). The solid line is the best fit for Eq. \ref{WLF}. The aPP data (red, circles)
is shown below, along the results from 2D $^2$H exchange (blue, diamonds) and $^{13}$C relaxometry experiments \cite{Zemke}. The solid line
is a plot of the WLF function with $\tau_c(T_g) = 100$~K, $C_1 = 14.5$ and $C_2 = 30$~K, given in \cite{Zemke}.}
\label{fig:arrhenius}
\end{figure}

To support the $\langle\tau_{c}\rangle$ values obtained for the PIB, we compare our results with a dataset from the literature\cite{WhitePIB}.
Such data was measured using CODEX experiments, supposing an Arhenius activation function. We compare the results of such experiments with the  
ones using the DF-MSE in Fig. \ref{fig:arrhenius} (a). The nonlinear behavior typical of the WLF (or VTFH) activation function is clearly observed. 
The fit by the WLF activation function 
gives $\tau_c(T_g) = 90 \pm 10$~K, $C_1 = 12.3 \pm 0.1$ and $C_2 = 51 \pm 0.1$~K. Its worth mentioning that the previous measurements 
were performed using exchange NMR methods, which are sensitive to slower motions, i.e., longer correlation times (100 $\mu$s to 1 ms). 
Thus, they probe $\tau_c$ values at lower temperatures, meaning our approach offers an extension of the dynamic window attainable by 
those previous measurements. The complementarity of the two techniques is remarkable.

To validate the aPP data, we used the same WLF parameters as Zemke \emph{et al}\cite{Zemke} and make a comparison with their $^{13}$C 
NMR relaxation and $^{2}$H NMR exchange data, as shown in Fig. \ref{fig:arrhenius} (b). The parameters are $\tau_c(T_g) = 100$~K, 
$C_1 = 14.5$ and $C_2 = 30$~K. We observe a fair agreement among the different data sets and the WLF function, indicating a consistency 
between the DF-MSE technique with both $^{13}$C relaxometry and 2D $^2$H exchange NMR measurements. Our approach complements the two 
latter techniques as it is able to detect motions with correlation times in the 10$^{-4}$ to 10$^{-6}$~s range, highlighted by 
the temperature range between 280 and 300 K of Fig. \ref{fig:arrhenius} (b).

\section{Discussion and conclusions}

We proposed a simple method to study molecular motions using TD-NMR, based on the Goldman-Shen dipolar filter and in the Anderson-Weiss approximation. The experiment can be used to obtain the onset temperatures of molecular motions in the kHz-MHz frequency scale, providing an estimation of the temperatures where local or global molecular relations set in. It is also possible to obtain a reliable estimation of the motion parameters, such as, correlation times and their distributions as a function of temperatures, allowing to extract the activation functions of the molecular motions. The estimation of the distribution of correlation times are done using a using a Tykhonov regularization scheme. A drawback of such procedure is the need of a clean dataset to avoid numerical artefacts in the reconstructed distribution. Moreover, if the signal is long enough to have filter times in the time scale of T$_1$, the recovery of magnetization introduces a phase distortion in the signal, resulting in a false broadening of the distribution of correlation times. To avoid big distortions due to T$_1$ effects, the maximum filter time is limited to T$_1$.

Our approach has some advantages when compared to some NMR techniques often employed to probe 
molecular motions. It is independent on particular choices of activation function \cite{WhitePIB, White2, refMSEmotion} and the shape of the distribution of correlation times. Furthermore, as the measurements are carried out on the naturally abundant $^1$H nuclear spins there is no need to label the samples. This makes the experiments simpler and less expensive, due to the high signal-to-noise ratio. Finally, the DF-MSE sequence allows a very fast acquisition, taking few minutes to obtain one data point and offering a fairly wide dynamic window to probe molecular motions.  

\section{Acknowledgements}

The authors gratefully acknowledges financial support from Fundação de Amparo à Pesquisa do Estado de São Paulo (FAPESP), grant numbers  	
[2009/18354-8] and [2008/11675-0], and Conselho Nacional de Desenvolvimento Científico e Tecnológico (CNPq), grant numbers [312852/2014-2], [131489/2014-3], 
[401454/2014-2] and [300121/2015-6]. E.R.dA thanks Prof. Kay Saalwächter for useful discussions.


\begin{thebibliography}{99}
\bibitem{HernandezGluten} V. M. Hernandez-Izquierdo, D. S. Reid, T. H. Mchugh, J. D.  Berrios,and J. M. Krochta, J. M., J. Food Sci. , 73, 169 (2008).
\bibitem{FariaPFO} G. C. Faria, E. R. deAzevedo, H. von Seggern, Macromolecules , 46, 7865 (2013).
\bibitem{fariajpcb2009} G. C. Faria, T. S. Plivelic, R. F. Cossiello, A. A. Souza, T.D.Z. Atvars, I. L. Torriani, E. R. deAzevedo, J. Phys. Chem. B, 113, 11403 (2009).
\bibitem{HansenAdvMat} K. Do, Q. Saleem, M. K. Ravva, F. Cruciani, Z. P. Kan, J. Wolf, M. R. Hansen, P. M. Beaujuge, J. L. Bredas,Adv. Mater., 28, 8197 (2016).
\bibitem{Kurz} R. Kurz, A. Achilles, W. Chen, M. Schafer, A. Seidlitz, Y. Golitsyn, J. Kressler, W. G. Paul,G. Hempel, T. Miyoshi, T. Thurn-Albrecht, K. Saalwachter, K., Macromolecules, 50, 3891 (2017).
\bibitem{HongJPCB} T. Wang,H. Jo, W. F. H. DeGrado, M. Hong , J. Phys. Chem. B, 119, 4552 (2015).
\bibitem{DupreeNature} T. J. Simmons, J. C. Mortimer, O. D. Bernardinelli, A. C. Poppler, S. P. Brown, E. R. deAzevedo, R. Dupree, P. Dupree, Nature Comm., 7, 13902 (2016).
\bibitem{andronis1998} V. Andronis, G. Zografi, Pharmaceutical Research, 15, 835 (1998).
\bibitem{deAzevedoPNMRS} E. R. deAzevedo, T. J Bonagamba and D. Reichert, D., Prog. Nucl. Magn. Reson. Spect., 47, 137 (2005).
\bibitem{SpiessAnniversary} H. W. Spiess, Macromolecules, 50, 1761 (2017).
\bibitem{HansenRev} M. H. Hansen, Chem. Rev., 116, 1272 (2016).
\bibitem{Krushelnitsky} A. Krushelnitsky, D. Reichert, K. Saalwachter, Accounts chem. Res., 46, 2028 (2013).
\bibitem{deAzevedo2009}  E. R. deAzevedo, W.G. Hu, T. J. bonagamba, K. Schmidt-Rohr, J. Am. chem. Soc., 121, 8411 (1999).
\bibitem{faske2008} S. Faske, H. Eckert, M. Vogel, Phys. Rev. B, 77, 104301 (2008).
\bibitem{deazevedo2000} E. R. deAzevedo, S. B. Kennedy, M. Hong, Chem. Phys. Lett., 321, 43 (2000).
\bibitem{palmer2001} A. G. Palmer, C. D. Kroenke, J. P. Loria, Methods Enzym., 339, 204 (2001).
\bibitem{lange2011} F. Lange, K. Schwenke, M. Kurakazu, Y. Akagi, U. I. Chung, M. Lane, J. U.
Sommer, T. Sakai, K. Saalwachter, Macromolecules, 44, 9666 (2011).
\bibitem{rothwelljcp1981} W. P. Rothwell, J. S. Waugh, J. Chem. Phys., 74, 2721 (1981).
\bibitem{cobopccp2009} M. F. Cobo, K. Malinakova, D. Reichert, K. Saalwachter, E. R. deAzevedo, Phys. Chem.
Chem. Phys., 11, 7036 (2009).
\bibitem{Schaler2013} K. Schaler, A. Achilles, R. Barenwald, C. Hackel, K. Saalwachter, Macromolecules, 46, 7818 (2013).
\bibitem{Saalwachter2012} K. Saalwachter, Macromolecules, 85, 350 (2012).
\bibitem{Papon2011} A. Papon, K. Saalwachter, K. Schaler, L. Guy, F. Lequeux, H. Montes, Macromolecules, 44, 913 (2011).
\bibitem{refMSEmotion} S. Sturniolo and K. Saalwätcher, Chem. Phys. Lett. 516, 106 (2011).
\bibitem{GS1} M. Goldman and L. Shen, Phys. Rev. 144, 321 (1966). 
\bibitem{Matsui} S. Matsui, J. of Magn. Res. 98, 618 (1992).
\bibitem{AWeiss} P. W. Anderson and P. R. Weiss, Rev. of Mod. Phys. 25, 269 (1953).
\bibitem{WhitePIB} M. Wachowicz and J. L. White, Macromolecules 40,  5433 (2007).
\bibitem{Zemke} K. Zemke, K. Schmidt-Rohr and H. U. Spiess, Acta Polymer., 45, 148 (1994).
\bibitem{SpiessDF} F. Mellinger , M. Wilhelm and H. W. Spiess, Macromolecules, 32, 4686 (1999).
\bibitem{MAPE} D. E. Demco, A. Johansson, and J. T. Tegenfeldt, Solid State Nucl. Mag. Res. 4, 13 (1995).
\bibitem{SpinDiffBlumich_KaySchaeller} K. Schäler, M. Roos, P. Micke, Y. Golitsyn, A. Seidlitz, T. Thurn-Albrecht, H. Schneider, 
G. Hempel, K. Saalwächter. Solid State Nucl. Magn. Reson. 72, 50-63 (2015).
\bibitem{Packer} K. J. Packer and  J. M. Pope, J. of Magn. Res. 55, 378 (1983).
\bibitem{Wokaun} A. Wokaun and R. R. Ernst, Chem. Phys. Lett. 52, 407 (1977). 
\bibitem{HanhEcho_CPMG} E. L. Hahn Phys. Rev. 80, 580 (1950); H. Y. Carr, and e. M. Purcell, Phys. Rev. 94, 630 (1954); 
S. Meiboom, and D. Gill, Rev. Sci. Instrum., 29, 688 (1958).
\bibitem{Boyd1} R. H. Boyd, Polymer 26, 323 (1985).
\bibitem{Boyd2} R. H. Boyd, Polymer 26, 1123 (1985).
\bibitem{KayKerstin} R. Bärenwald, Y. Champouret, K. Saalwächter, K. Schäler, J. Phys. Chem. B 116, 13089 (2012).
\bibitem{activationfunctions} G. Tammann, W. Hesse, Z. Anorg. Allg. Chem. 156. 245  (1926). D. Davidson, R. Cole, J. Chem. Phys. 19, 1484 (1951). 
M. L. Williams, R. F. Landel, and J. D. Ferry, J. Amer. Chem. Soc. 77, 3701 (1955).
\bibitem{VanVleck} J. H. van Vleck,Phys. Rev., 74, 1168 (1948).
\bibitem{Hirschinger1} J. Hirschinger, S. S. Nucl. Magn. Reson., 34,210 (2008).
\bibitem{Hirschinger2} J. Hirschinger, Concepts Magn. Reson., 28A, 307 (2006).
\bibitem{Kimmich1} R. Kimmich. NMR: tomography, diffusometry, relaxometry. Berlin: Springer, 1997. 
\bibitem{Kimmich2} R. Kimmich. Principles of soft-matter dynamics. Berlin: Springer, 2012.
\bibitem{Reichert} D. Reichert, K. Saalwächter. Dipolar coupling: Molecular-level mobility. eMagRes, 2007.
\bibitem{log-gaussianSpiess1} S. Wefing, H. W. Spiess, J. Chem. Phys., 89, 1219 (1988).
\bibitem{log-gaussianSpiess2} S. Wefing, S. Kaufmann, H. W. Spiess, J. Chem. Phys., 89, 1234 (1988).
\bibitem{White2} M. Wachowicz, L. Gill, J. E. Wolak, J. L. White, Macromolecules 41, 2832 (2008).
\bibitem{CODEX} E. R. DeAzevedo, W. G. Hu, T. J. Bonagamba, K. Schmidt-Rohr, J. of Am. Chem. Soc. 121, 8411 (1999).
\bibitem{PUREXinterm} E. R. DeAzevedo, T. J. Bonagamba, K. Schmidt-Rohr, J. Mag. Res. 142, 86 (2000).
\bibitem{Dipshift1} E. R. deAzevedo, K. Saalwachter, O. Pascui, A. A. de Souza, T. J. Bonagamba, and D. Reichert. 
J. Chem. Phys. 128, 104505 (2008).
\bibitem{Tikhonov} S. W. Provencher, Comput. Phys. Commun. 27, 213 (1982). G. C. Borgia, R. J. S. Brown, and P. Fantazzi, J. Magn. 
Res.  132, 65 (1998). 
\bibitem{Abragam} A. Abragam, \textit{The Principles of Nuclear Magnetism}. Clarendon Press (1961).


\end{thebibliography}
\end{document}